\documentclass{article}

\usepackage[preprint]{neurips_2024}

\usepackage[utf8]{inputenc} 
\usepackage[T1]{fontenc}    
\usepackage{hyperref}       
\usepackage{url}            
\usepackage{booktabs}       
\usepackage{amsfonts}       
\usepackage{nicefrac}       
\usepackage{microtype}      
\usepackage{xcolor}         
\usepackage{graphicx}
\usepackage{listings}

\lstnewenvironment{verbatim}{\lstset{breaklines,basicstyle=\ttfamily}}{}

\title{PEAR: A Robust and Flexible Automation Framework for Ptychography Enabled by Multiple Large Language Model Agents}

\author{
  Xiangyu Yin \\
    Argonne National Laboratory \\
    Lemont, IL \\
  \texttt{xyin@anl.gov} \\
  \And
  Chuqiao Shi \\
  Rice University \\
  Houston, Tx \\
  \texttt{cs109@rice.edu} \\
  \And
  Yimo Han \\
  Rice University \\
  Houston, Tx \\
  \texttt{yh76@rice.edu} \\
  \And
    Yi Jiang \\
    Argonne National Laboratory \\
    Lemont, IL \\
  \texttt{yjiang@anl.gov} \\
}

\begin{document}
\section*{GOVERNMENT LICENSE}

The submitted manuscript has been created by UChicago Argonne, LLC, Operator of Argonne National Laboratory (“Argonne”). Argonne, a U.S. Department of Energy Office of Science laboratory, is operated under Contract No. DE-AC02-06CH11357. The U.S. Government retains for itself, and others acting on its behalf, a paid-up nonexclusive, irrevocable worldwide license in said article to reproduce, prepare derivative works, distribute copies to the public, and perform publicly and display publicly, by or on behalf of the Government. The Department of Energy will provide public access to these results of federally sponsored research in accordance with the DOE Public Access Plan. \href{http://energy.gov/downloads/doe-public-access-plan}{http://energy.gov/downloads/doe-public-access-plan}

\newpage

\maketitle

\begin{abstract}
   Ptychography is an advanced computational imaging technique in X-ray and electron microscopy. It has been widely adopted across scientific research fields, including physics, chemistry, biology, and materials science, as well as in industrial applications such as semiconductor characterization. In practice, obtaining high-quality ptychographic images requires simultaneous optimization of numerous experimental and algorithmic parameters. Traditionally, parameter selection often relies on trial and error, leading to low-throughput workflows and potential human bias. In this work, we develop the “Ptychographic Experiment and Analysis Robot” (PEAR), a framework that leverages large language models (LLMs) to automate data analysis in ptychography. To ensure high robustness and accuracy, PEAR employs multiple LLM agents for tasks including knowledge retrieval, code generation, parameter recommendation, and image reasoning. Our study demonstrates that PEAR’s multi-agent design significantly improves the workflow success rate, even with smaller open-weight models such as LLaMA 3.1 8B. PEAR also supports various automation levels and is designed to work with customized local knowledge bases, ensuring flexibility and adaptability across different research environments.
\end{abstract}

\section{Introduction}
\label{intro}
Ptychography\citep{hoppe1969beugung} is a powerful computational imaging technique widely used in optical, X-ray, and electron microscopy. Unlike conventional imaging methods, whose spatial resolution is fundamentally limited by focusing optics, modern ptychography employs optimization-based algorithms to directly reconstruct sample structures from a series of scanning diffraction patterns\citep{rodenburg2004phase}. This approach overcomes physical limitations imposed by the lenses, enabling much higher resolution under similar conditions. Currently, electron ptychography holds the world record for the highest spatial resolution among all microscopy techniques. To date, ptychography has enabled numerous scientific applications, ranging from in-situ studies of battery particles\citep{liu2022conformal} and 3D imaging of integrated circuits\citep{aidukas2024high} to single oxygen vacancy detection in superconducting materials\citep{dong2024visualization}. 

Despite its popularity, achieving high-resolution ptychography remains challenging due to various parameters that significantly impact reconstruction quality and require careful adjustment for different datasets or applications. For example, physical parameters describing processes such as noise generation, partial coherence, and multiple scattering can be modeled in the optimization problem. Choosing appropriate values for these parameters to account for various practical errors is paramount to reaching solutions that closely represent the real object. Other parameters, including the number of iterations, update step size, and initial solutions, also impact reconstructions by controlling the convergence process. Given the virtually infinite possibilities and complex trade-offs between parameters, optimizing all variables by exhaustive search is practically impossible. In practical applications\citep{jiang2018electron, song2019atomic, chen2020dataset}, researchers often rely on their own experiences and knowledge to choose a limited set of parameters through trial and error. This process reduces the overall throughput and potentially introduces biases to scientific conclusions drawn from the results. Therefore, there is a critical need to develop workflows that automatically optimize parameters for unknown datasets with minimal human input. 

Previously, black-box optimization techniques such as Bayesian optimization with Gaussian processes (BO-GP) have been developed for parameter tuning tasks in ptychography\citep{cao2022automatic}. BO-GP workflows make minimal assumptions about the input data or reconstruction algorithms, iteratively searching for the global optima by automatically balancing exploration and exploitation. This approach has found great success in calibrating experimental parameters that are difficult to measure by instruments\citep{yoon2024improving}, and in designing complex experiments involving multiple trade-offs between parameters\citep{cao2022automatic}. However, BO-GP requires a large number of reconstructions to construct good surrogate models that approximate the objective function, which can significantly prolong processing time and increase the computational resource demands compared with experienced researchers whose existing knowledge can guide them obtaining useful results within a few reconstructions.

The recent advancement of foundation models has propelled intelligence (AI) to a new era characterized by billion-parameter models and multifunctional capabilities. Recently, Large Language Models (LLMs), such as those popularized by OpenAI’s ChatGPT, have gained significant attention due to their “intelligent” capabilities, including comprehensive knowledge, common sense, and basic logic. With extensive training datasets, these foundation models can be easily adapted to automate various tasks including literature search, reasoning, code generation, which are key components in many scientific field. \citet{boiko2023autonomous} developed an LLM-powered system named Coscientist to autonomously design, plan and perform several organic chemistry experiments. \citet{m2024augmenting} introduced ChemCrow, an LLM chemistry agent that can perform tasks across organic synthesis, drug discovery, and materials design by integrating 18 expert-designed tools. \citet{calms} proposed Context-Aware Language Model for Science (CALMS) to assist scientists with instrument operations and complex experimentation in the scientific user facilities.

In this work, we developed the "Ptychographic Experiment and Analysis Robot" (PEAR), a workflow designed to assist researchers with parameter tuning in ptychographic reconstructions. PEAR mimics human experts’ thought process based on domain knowledge embedded within multiple LLM agents to automate and robustly perform tasks such as information collection, code generation, knowledge retrieval, and image reasoning. We designed various levels of automation, ranging from human-in-the-loop to fully automated closed-loop, to support a broad range of users, from novice researchers to experienced experts, tailored to their customized knowledge bases. These features make PEAR highly adaptable, allowing it to be easily extended to other computational imaging techniques and beyond.

\section{Method}
\label{method}
In this section, we introduce the main design features of PEAR, which are critical for the workflow to consistently produce accurate parameters and high-quality ptychographic reconstructions.

\subsection{Custom Knowledge Base}
Effectively utilizing domain knowledge is crucial in applying LLMs for scientific applications to ensure that their outcomes are both scientifically accurate and tailored to specific scientific tasks at hand. In PEAR, we decouple the workflow from the knowledge bases. The workflow itself contains no specific information about ptychography, while the knowledge bases serve as dedicated repositories for problem- and user-specific information, such as experiment settings, example scripts, related documents, and images. By leveraging a custom knowledge base, the PEAR workflow can make informed decisions and provide guidance tailored to specific tasks in ptychography reconstruction.

This separation of the workflow from the knowledge bases allows for dynamic updating or modification of either component without affecting the other. For example, as new ptychographic techniques or parameters are developed, the knowledge base can be expanded without altering the core workflow logic. Similarly, improvements can be made to the LLMs and the overall agentic workflows without modifying the knowledge bases. This separation is particularly valuable in practical scientific research, as it allows for customization and ensures information privacy for different research groups.

\begin{figure}
  \centering
    \includegraphics[width=\textwidth]{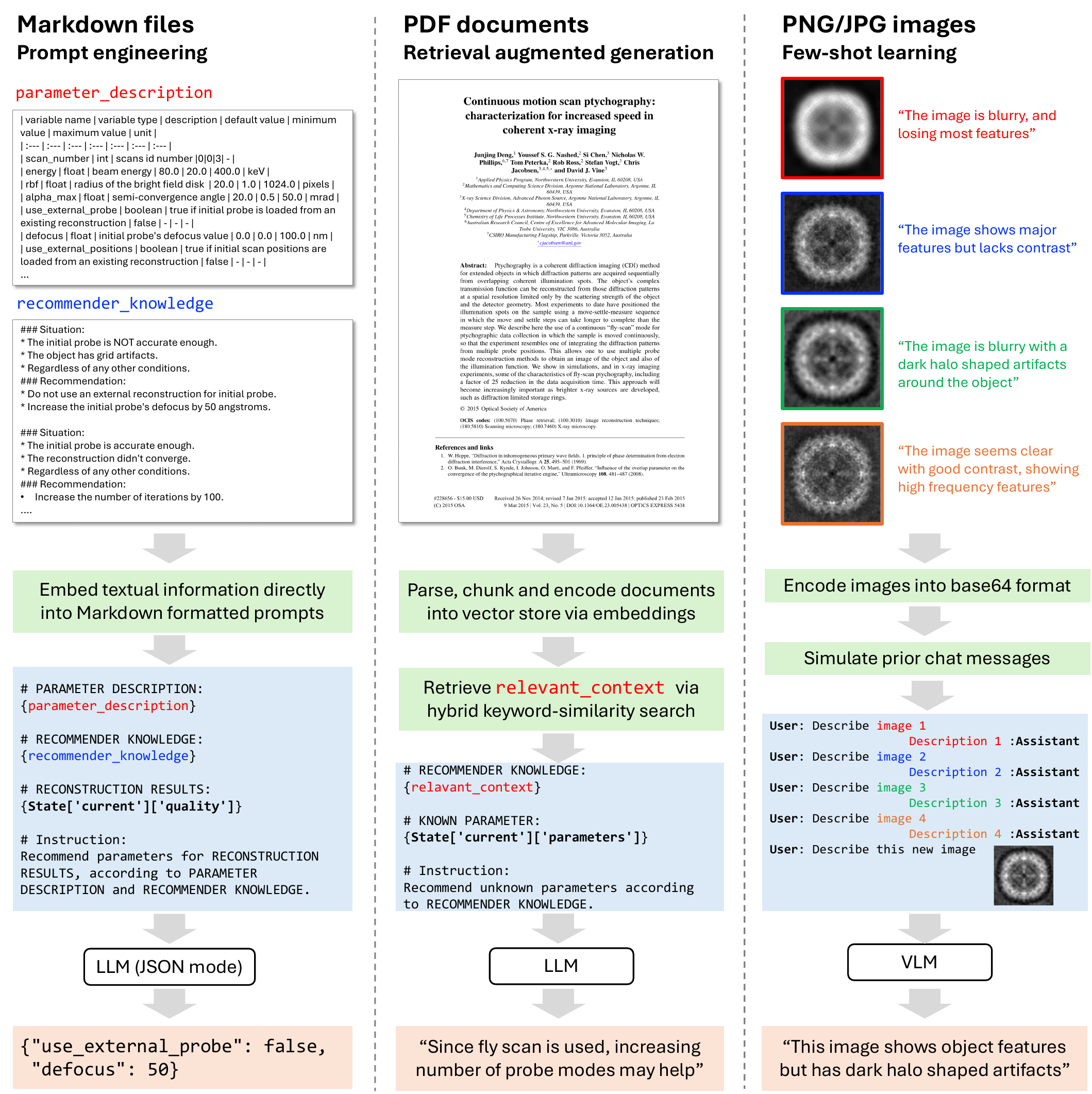}
  \caption{Supported knowledge files in PEAR. (\textbf{Left}) Markdown files are used for structured, text-based information such as parameter descriptions or step-by-step protocols. In PEAR, these files can be directly embedded into prompts and formatted using prompt engineering techniques. For example, parameter descriptions enable LLMs to understand the nuances of various ptychographic parameters, their acceptable ranges, and their impacts on the reconstruction process. The knowledge base also includes expert guidance on parameter selection and optimization strategies, allowing LLMs to make informed recommendations based on specific reconstruction results and conditions. JSON formatting ensures that model outputs are consistent and machine-readable, facilitating seamless integration with other components of the PEAR system. (\textbf{Middle}) PDF documents, including scientific papers, user manuals, and technical reports, provide richer, more detailed information. PEAR processes these documents using Retrieval Augmented Generation techniques, which involve parsing the content, chunking it into manageable sections, and encoding these chunks into a vector store using embeddings. When required, PEAR efficiently retrieves relevant information through hybrid keyword-similarity searches, ensuring that the most pertinent content informs its outputs. (\textbf{Right}) Images (PNG/JPG) are critical for representing ptychographic data and results. In PEAR, images are encoded in base64 format, making them compatible with most vision-language models. Few-shot learning techniques enable PEAR to understand and describe new images based on previously seen examples. This approach simulates prior interactions where the system describes various ptychographic images, providing a foundation for interpreting new, unseen data. To process and understand image content, PEAR incorporates Visual Language Models that analyze ptychographic images and generate descriptive outputs, which inform parameter recommendations and quality assessments.
}
  \label{kb}
\end{figure}

In PEAR’s knowledge base, we limit the types of knowledge sources to Markdown files, PDF documents, and PNG/JPG images, which encompass most of the information types relevant to ptychography. Regulating the types of knowledge sources ensures effective and reliable storage, retrieval, and utilization of information using common LLM techniques. Additionally, these source types are human-readable and manageable, making it possible to trace back LLM outputs, thereby enhancing user interpretability and trust. Figure~\ref{kb} illustrates three different file formats in PEAR's knowledge base and how they facilitate various sub-tasks in the data analysis workflow.

\subsection{Human-in-the-Loop (HITL)}
Despite their impressive capabilities, Large Language Models (LLMs) can sometimes produce errors or inconsistencies, particularly when dealing with complex scientific concepts. Human-in-the-Loop (HITL) serves as a critical quality control mechanism, allowing human experts to intervene and correct subtle errors that might otherwise go unnoticed by LLMs. Scientific research often involves navigating conflicting information, scenarios where standard protocols do not apply, and outcomes that are inherently uncertain. HITL plays a vital role in managing these edge cases and preventing catastrophic failures.

In PEAR, the HITL approach is integrated at various stages of the workflow, performing several essential functions:

\begin{itemize}
\item Providing Contextual Information: Human users can offer context-specific details, such as experimental setups or prior knowledge about the sample that are not captured in the general knowledge base. This input is facilitated through responses to targeted questions generated by LLM agents, ensuring that the workflow remains grounded in realistic experimental conditions.
\item Validation and Error Correction: Human users can verify LLM-generated values and recommendations, ensuring alignment with scientific principles and experimental constraints. Experienced researchers can identify and correct errors early in the workflow, preventing the propagation of inaccuracies and maintaining the integrity of the scientific analysis.
\item Assessment and Feedback: Experts can evaluate ptychographic reconstructions, enabling an iterative process where their assessments guide LLM agents in refining the workflow. Feedback from human experts is logged and can be used to enhance the knowledge base over time, contributing to continuous improvement and adaptation of the system. 
\end{itemize}

\subsection{Multi-agent Workflow}
\label{test}
\begin{figure}
  \centering
    \includegraphics[width=\textwidth]{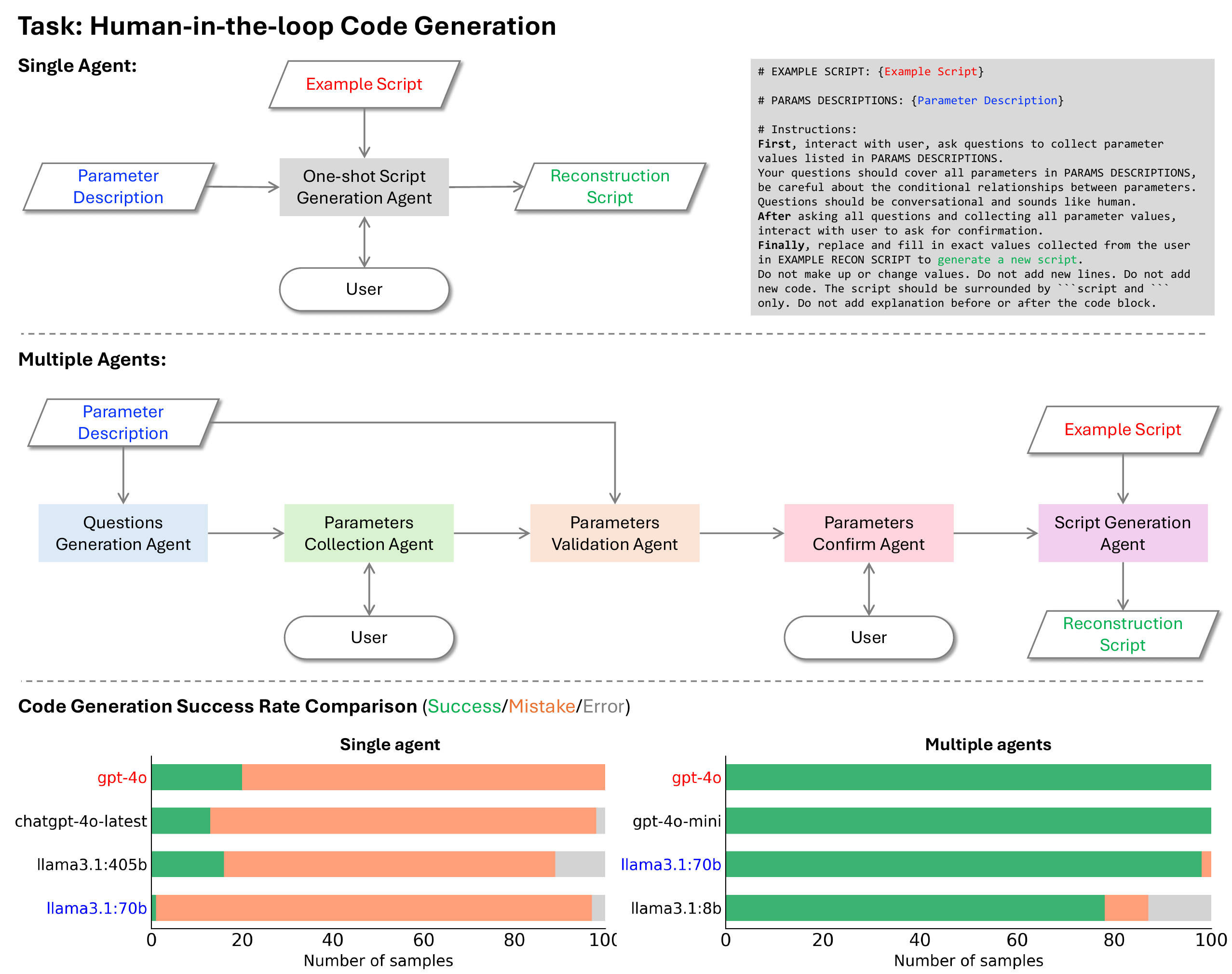}
  \caption{Comparison of single-Agent vs. multi-Agent Design for Code Generation in PEAR. The single-agent workflow uses a single prompt to collect parameter values from users and generate a reconstruction script based on their inputs. The multi-agent workflow breaks the task into simpler sub-tasks, each handled by specialized agents. The Question Generation Agent formulates questions to gather necessary information from the user. The Parameter Collection Agent processes user responses and extracts relevant parameter values. The Parameter Validation Agent checks these values for consistency and compliance with domain-specific constraints. The Parameter Confirmation Agent interacts with the user to confirm the collected parameters before script generation. Finally, the Script Generation Agent uses the validated parameters to generate the complete reconstruction script, enhancing accuracy and reducing errors compared to the single-agent approach.}
  \label{multi}
\end{figure}

To further enhance the reliability and effectiveness of LLM-powered scientific workflows, particularly in complex tasks like ptychography data analysis, PEAR employs multiple LLM agents for each sub-task instead of relying on a single agent. This design choice enables PEAR to achieve nearly 100\% accuracy in its outputs. One of the core tasks in PEAR is the HITL code generation task, where experimental information is gathered from users to generate a corresponding reconstruction script. We use this task as an example to compare single-agent and multi-agent approaches, illustrating the benefits of the multi-agent design.

In a single-agent system, one LLM agent manages the entire process, including user interaction, parameter collection, and script generation. Although simpler, this approach is more susceptible to errors and inconsistencies, especially when handling complex parameter relationships and domain-specific constraints. In contrast, the multi-agent approach in PEAR distributes these tasks among specialized agents, each responsible for a specific aspect of the workflow.

As shown in Figure~\ref{multi}, we conducted a computational experiment comparing the single-agent and multi-agent approaches. We generated 100 sets of hypothetical ptychography problems and produced reconstruction scripts for each condition using both approaches. To efficiently handle the large number of tests, we prompted an advanced LLM model to simulate human responses in a HITL setup. Both the single-agent and multi-agent approaches were tested using various LLMs, including open-weight models like LLaMA 3.1 and closed-source models like ChatGPT. Performance was evaluated by comparing the generated scripts against expert-curated ground truth scripts, classifying the outcomes into three categories: success, mistake, and error. The success category includes scripts that are correct and fully functional. The mistake category consists of runnable scripts with incorrect parameter values, while the error category includes scripts with syntax errors that prevent reconstruction. Detailed experiment setups and prompts are provided in the Appendix.

The results demonstrate that the multi-agent approach significantly enhances the success rate and robustness of the code generation process. This increased accuracy allows PEAR to utilize open-weight models with fewer parameters compared to popular models like GPT-4, offering significant advantages for scientific workflows. Open-weight models can be fine-tuned or trained on task-specific datasets, enhancing customization. Furthermore, these models can be deployed locally, ensuring reproducibility and data security using existing infrastructure.

\subsection{Automation Levels}
Inspired by the self-driving car industry, PEAR offers different levels of automation to enhance the framework's versatility and user-friendliness. These levels accommodate users with varying degrees of expertise and familiarity with ptychography and autonomous systems, allowing both novice and experienced users to effectively utilize the platform. Newcomers to ptychography can begin with lower levels of automation and gradually advance to more sophisticated features as they gain confidence and understanding in data analysis. Additionally, as users interact with PEAR at different automation levels, the system automatically tracks and saves valuable data on user behavior, decision-making processes, and problem-solving strategies. This multi-modal data, comprising both text and images, is rare in the scientific community and is crucial for developing improved knowledge bases and more intelligent LLM agents.

\begin{figure}
  \centering
    \includegraphics[width=\textwidth]{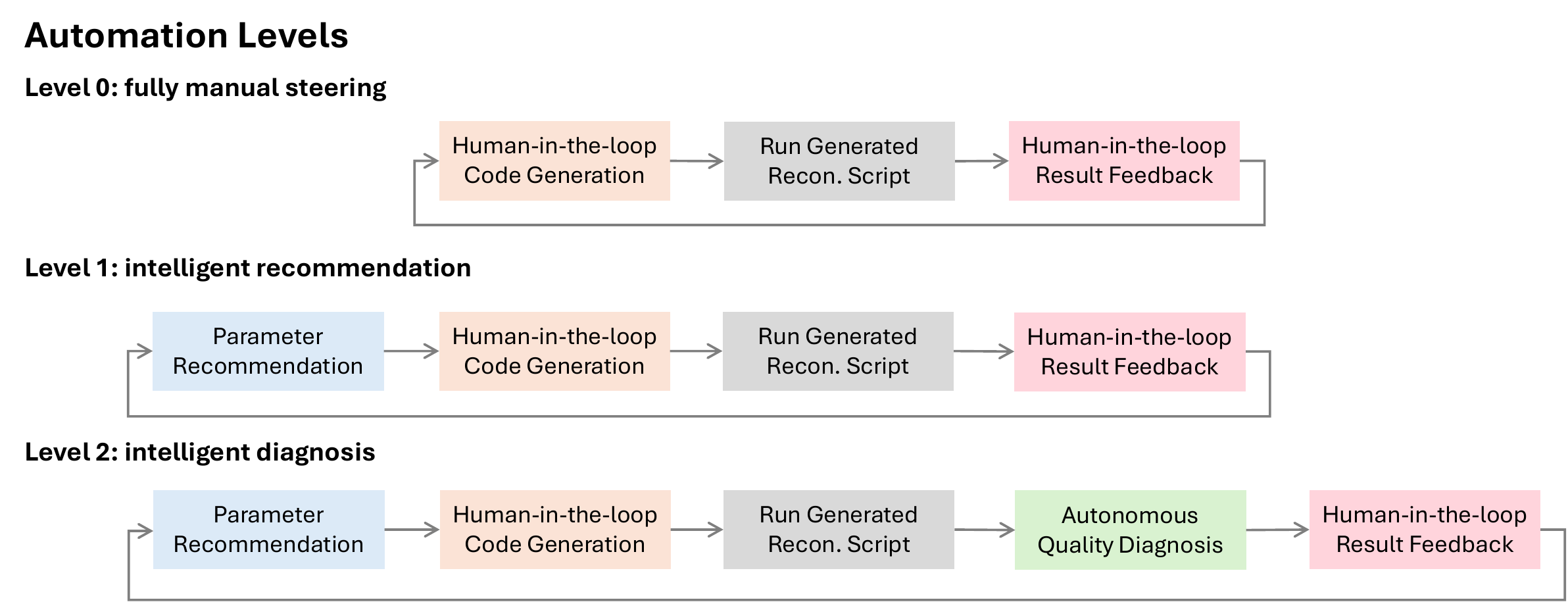}
  \caption{Levels of automation in PEAR. At \textbf{level 0}, human users maintain full control over the reconstruction process. They manually specify input parameters and customize the code according to their specific needs. PEAR then generates a reconstruction script based on these inputs. After reconstruction, users manually examine the results and decide on further steps or parameter adjustments. Even at this manual level, the system collects data on user inputs and decisions, which is valuable for refining future recommendations and enhancing automation capabilities. At \textbf{level 1}, users still control code generation, but the system provides intelligent suggestions based on the input data and experimental parameters. PEAR generates a reconstruction script that integrates both user inputs and AI-recommended parameters. After the reconstruction, PEAR offers recommendations for further parameter adjustments to improve quality based on user feedback. Users have the option to accept, modify, or reject these recommendations in subsequent iterations. At \textbf{level 2}, PEAR's diagnostic agents automatically assess the quality of the reconstruction, identifying potential issues such as artifacts or convergence problems. The AI provides a quality assessment that users can review, offering insights and feedback on the reconstruction quality and suggesting further actions or adjustments.}
  \label{level}
\end{figure}

Currently, PEAR implements three primary levels of automation, as shown in Figure~\ref{level}, each offering an increasing degree of AI assistance and autonomous operation:

\begin{itemize}
\item Level 0: Manual Workflow Steering: This level represents the most basic form of automation in PEAR, designed for users who prefer maximum control over the ptychographic reconstruction process or for scenarios requiring careful human oversight.

\item Level 1: AI-Assisted Recommendations: At this level, PEAR provides AI-generated recommendations based on the user’s knowledge base while maintaining significant user involvement. It is ideal for users who wish to leverage AI insights but still retain decision-making control.

\item Level 2: Automated Quality Assessment: The highest level currently implemented in PEAR, Level 2 incorporates automated quality assessment of reconstruction results. This level is particularly suited for experienced users or for handling large datasets where manual inspection of each result would be inefficient and time-consuming.
\end{itemize}

The multi-level design of PEAR is flexible and adaptable, allowing the framework to be easily modified for use in other computational imaging techniques and beyond.

\section{A Case Study in Electron Ptychography}
\label{case}
\begin{figure}
  \centering
    \includegraphics[width=\textwidth]{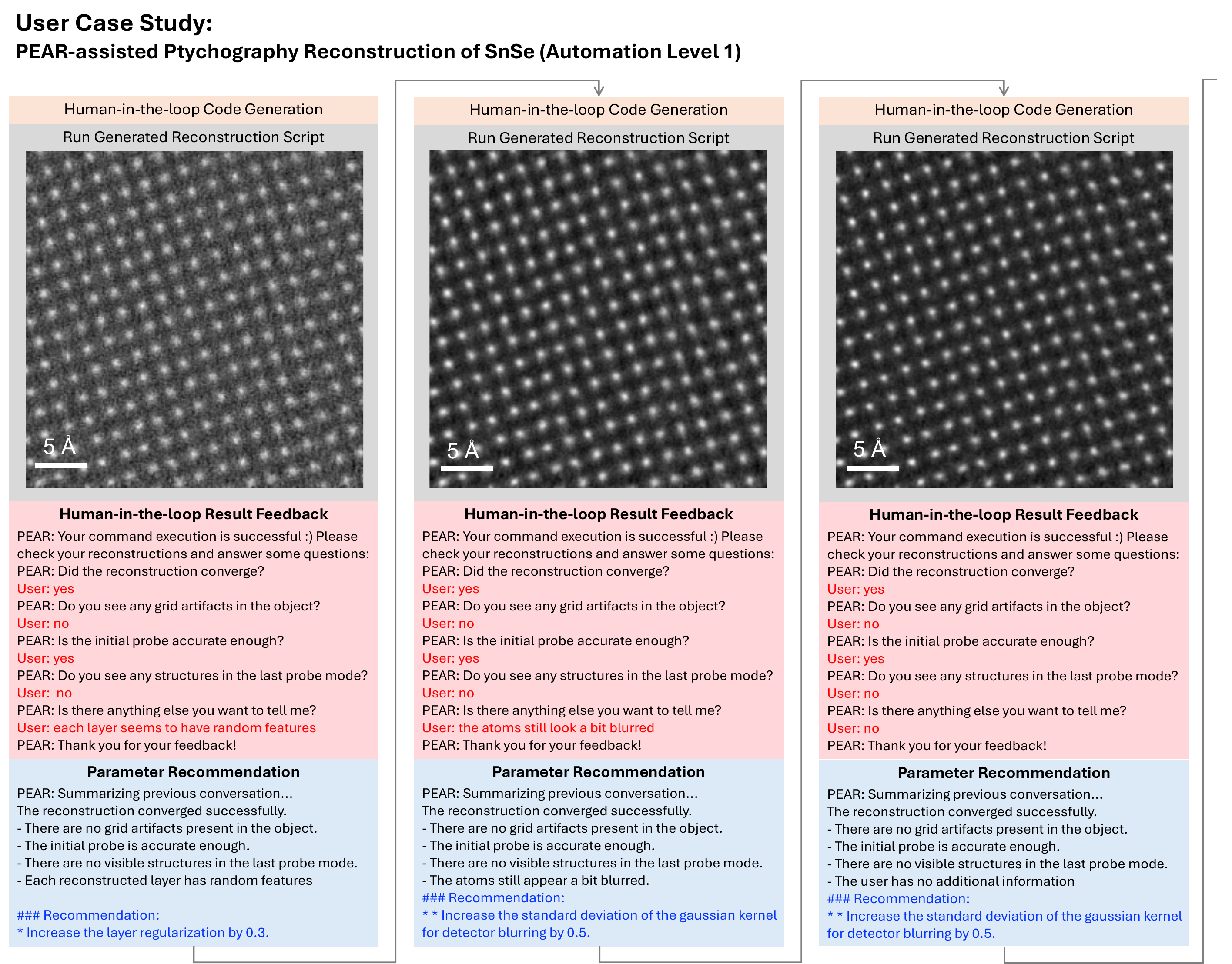}
  \caption{PEAR-assisted reconstructions of an electron ptychography dataset of SnSe. At automation Level 1, PEAR incorporates user feedback to recommend optimized parameters that enhance image quality. As a result, the reconstructed atomic structures become noticeably sharper and contain less noisy artifacts throughout the PEAR-guided workflow.}
  \label{study}
\end{figure}

To further demonstrate PEAR's workflow, we applied it to an experimental electron ptychography dataset of 2D Van der Waals SnSe, a material with structures that are crucial for understanding the electronic properties of ferroelectrics \citep{zhang2023ferroelectric}. Figure~\ref{study} shows selected ptychographic reconstructions along with the corresponding evaluations and recommendations recorded by PEAR. The complete log of the parameter tuning process is provided in the appendix, with some sensitive data redacted to protect user privacy.

At automation Level 1, PEAR leverages user feedback on the reconstruction quality to suggest adjustments to reconstruction parameters. In a few reconstructions, the reconstructed atomic structures become notably sharper and exhibit fewer noisy artifacts when using parameters recommended by PEAR's knowledge base. These optimized choices closely align with those made by experienced scientists. Without the knowledge about ptychography, conventional BO approaches could require tens of reconstructions to achieve similar quality\citep{cao2022automatic}. 

\section{Conclusions}
\label{conclusion}
In this work, we present PEAR, an automatic parameter-tuning workflow for ptychography powered by Large Language Models. To ensure more robust and accurate results, PEAR incorporates several key strategies, including custom knowledge bases, human-in-the-loop integration, and a multi-agent workflow. These features enable PEAR to achieve high success rates even when using smaller LLM models, such as LLaMA 3.1 8B. Despite the great improvement, we acknowledge that errors and hallucinations can still happen given the nature of current foundation model technologies. The different levels of automation provide flexibility and adaptability, allowing PEAR to cater to a wide range of users and extend automation capabilities to computational imaging techniques beyond ptychography. Looking ahead, the log files generated by PEAR users have the potential to serve as valuable training data for the development of future foundation models. 

\section{Acknowledgements}
\label{Acknowledgements}
This research used resources of the Advanced Photon Source, an U.S. Department of Energy (DOE) Office of Science User Facility operated for the DOE Office of Science by Argonne National Laboratory under Contract No. DE-AC02-06CH11357. X.Y. and Y.J. acknowledge funding support from Argonne LDRD 2023-0049. C.S. and Y.H. acknowledge NSF (FUSE-2329111 and CMMI–2239545), Welch Foundation (C-2065), and American Chemical Society Petroleum Research Fund (67236-DNI10).
\newpage
{
\small
\bibliographystyle{plainnat}
\bibliography{pear}
}


\appendix

\section{Computational Experiment Setup}
\begin{figure}
  \centering
    \includegraphics[width=\textwidth]{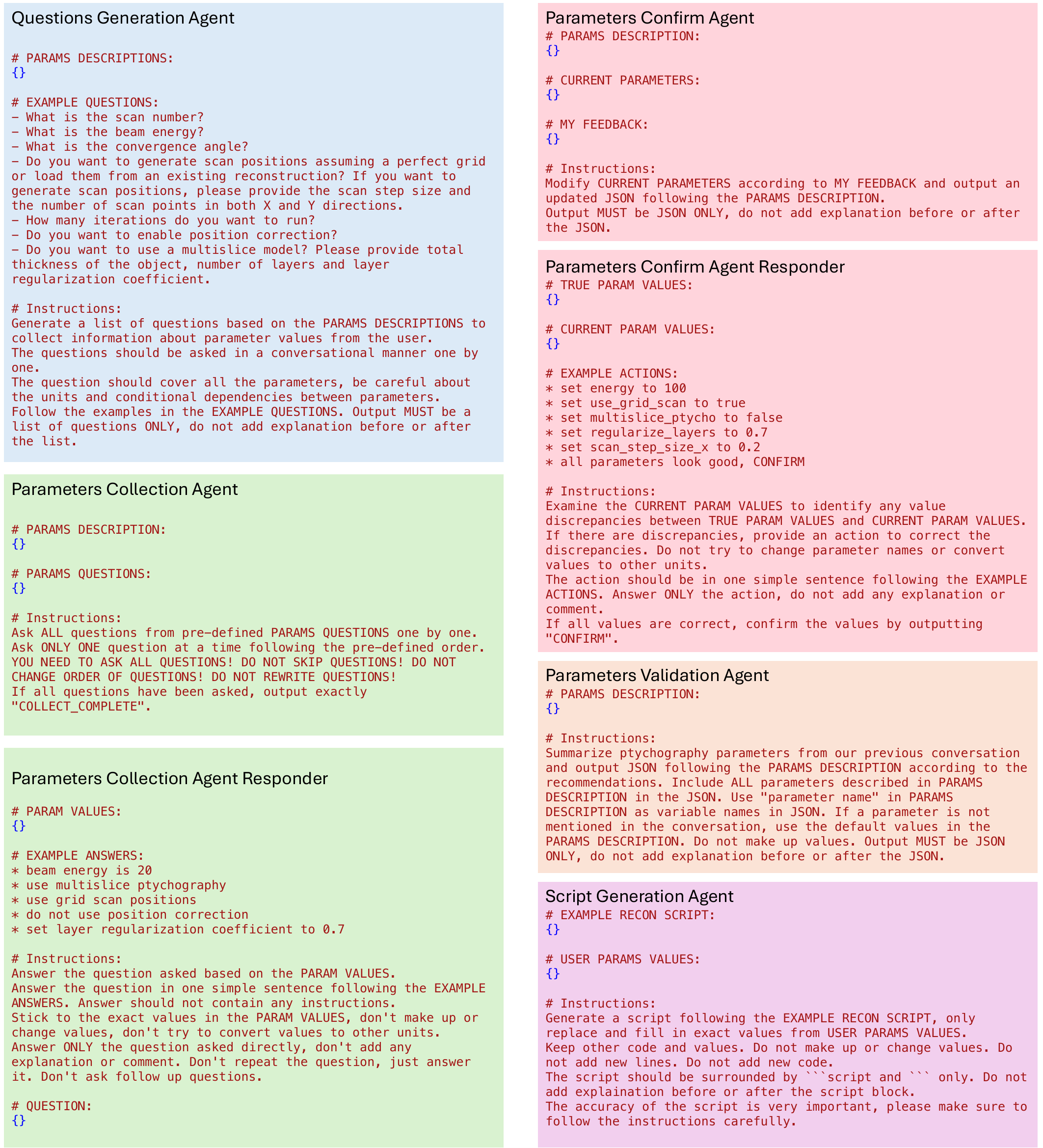}
  \caption{Prompts used in the computational experiment.}
  \label{prompts}
\end{figure}

In the computational experiment described in Section~\ref{test}, we used an advanced LLM to simulate human responses to the LLM agents' questions, enabling us to efficiently conduct a large number of tests. Specifically, we employed the GPT-4 model via Azure OpenAI Service to simulate human responses. For the LLM agents, we utilized a range of models, including LLaMA 3.1 (8b, 70b, 405b) through Together AI, and ChatGPT models (gpt-4o, chatgpt-4o-latest, gpt-4o-mini) via OpenAI. The exact prompts used in these experiments are detailed in Figure~\ref{prompts}.

The hypothetical experiment conditions and computational experiment logs are available in the supplementary materials. 

\newpage

\section{User Case Study Log}
\tiny
\begin{verbatim}
20240913-00:02:56 - Initializaing PEAR...
20240913-00:02:56 - ########## Settings ##########
20240913-00:02:56 - Automation level: 1
20240913-00:02:56 - LLM: gpt-4o-mini
20240913-00:02:56 - Knowledge base: neurips_demo
20240913-00:02:56 - Data's base directory: /Modified/to/Hide/User/Info/
20240913-00:02:56 - Script directory: /Modified/to/Hide/User/Info/ptycho
20240913-00:02:56 - User name: User
20240913-00:02:56 - Computer name: lamda
20240913-00:02:56 - Matlab directory: /usr/local/bin/matlab
20240913-00:02:56 - External reconstruction script: 
20240913-00:02:56 - ##############################
20240913-00:02:56 - PEAR: A driver file exists in the current working directory, it will be overwritten.
20240913-00:02:56 - PEAR: Driver file created at driver.m
20240913-00:02:56 - PEAR: Hello User. Thank you for letting me assist with your ptychographic reconstruction today.
20240913-00:02:56 - Agent: ParamsCollector
20240913-00:02:56 - PEAR: First I'd like to ask some questions about your data:
20240913-00:02:56 - PEAR: What is the scan number?
20240913-00:02:59 - User: 31
20240913-00:03:00 - PEAR: What is the beam energy (in keV)?
20240913-00:03:03 - User: 300
20240913-00:03:03 - PEAR: What is the radius of the bright field disk (in pixels)? 
20240913-00:03:14 - User: 23.25
20240913-00:03:14 - PEAR: What is the convergence angle (in mrad)?
20240913-00:03:17 - User: 25
20240913-00:03:18 - PEAR: What is the detector size (in # of pixels)?
20240913-00:03:19 - User: 128
20240913-00:03:20 - PEAR: Do you want to load the initial object from an existing reconstruction? If so, please provide the path to the reconstruction file.
20240913-00:03:22 - User: no
20240913-00:03:22 - PEAR: Do you want to generate the initial probe based on the ideal model or load it from an existing reconstruction? If use ideal model, please provide the value of defocus (in angstroms). If use existing probe, please provide the path to the reconstruction file.
20240913-00:03:44 - User: load an existing probe from /Modified/to/Hide/User/Info/Niter1000.mat
20240913-00:03:44 - PEAR: Is the initial probe accurate (similar size to the real probe)?
20240913-00:03:46 - User: yes
20240913-00:03:47 - PEAR: Do you want to generate scan positions assuming a perfect grid or load them from an existing reconstruction? If you want to generate scan positions, please provide the scan step size (in angstroms) and the number of scan points in both X and Y directions. If use existing reconstructions, please provide the path to the reconstruction file.
20240913-00:04:05 - User: 128 x 128, step size is 0.2546 A
20240913-00:04:06 - PEAR: How many mixed-state probe modes do you want to use?
20240913-00:04:07 - User: 3
20240913-00:04:07 - PEAR: Which GPU do you want to use for reconstruction?
20240913-00:04:09 - User: 1
20240913-00:04:10 - PEAR: How many diffraction patterns does the scan have in total?
20240913-00:04:14 - User: 128 x 128
20240913-00:04:15 - PEAR: Did the sample drift during scan?
20240913-00:04:17 - User: no
20240913-00:04:17 - PEAR: What's the estimated sample thickness (in angstroms)?
20240913-00:04:20 - User: 100
20240913-00:04:20 - PEAR: Thank you for answering my questions.
20240913-00:04:20 - Agent: ParamsRecommender
20240913-00:04:20 - PEAR: Next I will initialize some reconstruction parameters based on your inputs.
20240913-00:04:20 - Agent: _summarize_conversation
20240913-00:04:20 - PEAR: Summarizing previous conversation...
20240913-00:04:23 - - The scan number is 31.  
- The beam energy is 300 keV.  
- The radius of the bright field disk is 23.25 pixels.  
- The convergence angle is 25 mrad.  
- The detector size is 128 pixels.  
- The initial object will not be loaded from an existing reconstruction.  
- An existing probe will be loaded from the path provided.  
- The initial probe is accurate and similar in size to the real probe.  
- The scan positions will be generated assuming a perfect grid with a step size of 0.2546 angstrom and 128 x 128 scan points.  
- Three mixed-state probe modes will be used.  
- GPU 1 will be used for reconstruction.  
- The total number of diffraction patterns for the scan is 128 x 128.  
- The sample did not drift during the scan.  
- The estimated sample thickness is 100 angstrom.  
20240913-00:04:25 - PEAR: Based on the provided information and the recommendations, here are the suggested reconstruction parameters:

1. **Update Batch Size**: Set the update batch size to \( \lfloor \sqrt{128 \times 128} \rfloor = 128 \).

2. **Gaussian Kernel Standard Deviation for Detector Blurring**: Since the beam energy is 300 keV, set the standard deviation of the Gaussian kernel for detector blurring to **1**.

3. **Scan Position Correction**: Since the sample did not drift during the scan, set scan position correction to **false**.

4. **Multislice Ptychography**: Given that the estimated sample thickness is 100 angstrom, set multislice ptychography to **true** and the number of layers to **10** (1/10 of the sample thickness).

5. **Number of Iterations**: Since the initial probe is accurate, set the number of iterations to **50**.

These parameters should help optimize your ptychography reconstruction process.
20240913-00:04:25 - Agent: ParamsFormatter
20240913-00:04:28 - PEAR: Here are the current parameters:
20240913-00:04:28 - {
    "data_directory": "/Modified/to/Hide/User/Info/",
    "scan_number": 31,
    "beam_energy": 300,
    "radius_bright_field": 23.25,
    "convergence_angle": 25,
    "size_of_diffraction_patterns": 128,
    "use_external_object": false,
    "initial_object_path": "",
    "use_external_probe": true,
    "initial_probe_file": "/Modified/to/Hide/User/Info/Niter1000.mat",
    "defocus": 0,
    "use_external_positions": false,
    "initial_position_file": "",
    "grid_scan_positions": true,
    "scan_step_size_x": 0.2546,
    "scan_step_size_y": 0.2546,
    "number_scan_points_x": 128,
    "number_scan_points_y": 128,
    "number_of_iterations": 50,
    "update_batch_size": 128,
    "number_of_probe_modes": 3,
    "position_correction": false,
    "multislice_ptycho": true,
    "object_thickness": 100,
    "number_of_layers": 10,
    "layer_regularization_coefficient": 0,
    "diff_pattern_blur": 1,
    "gpu_id": 1
}
20240913-00:04:28 - ########## Reconstruction No.1: ##########
20240913-00:04:28 - Agent: ParamsConfirmer
20240913-00:04:28 - PEAR: Let me know any changes you would like to make. If it looks good to you, please say "LGTM".
20240913-00:04:51 - User: lgtm
20240913-00:04:51 - PEAR: Thank you for the feedback. I will create a reconstruction script using these parameters.
20240913-00:04:51 - Agent: ScriptGenerator
20240913-00:04:51 - PEAR: Generating a reconstruction script...
20240913-00:04:57 - PEAR: The reconstruction script has been generated at /Modified/to/Hide/User/Info/ptycho/neurips_demo_gpt4o_mini_l1_ptycho_script_1.m
20240913-00:04:57 - Agent: ScriptRunner
20240913-00:04:57 - PEAR: Running ptychographic reconstruction...
20240913-00:04:57 - PEAR: Executing script using /usr/local/bin/matlab on Modified-to-Hide-User-Info...
20240913-00:04:57 - PEAR: The MATLAB command used for reconstruction: /usr/local/bin/matlab -batch "driver('/Modified/to/Hide/User/Info/ptycho/neurips_demo_gpt4o_mini_l1_ptycho_script_1.m')" 
20240913-00:04:57 - PEAR: Executing command /usr/local/bin/matlab -batch "driver('/Modified/to/Hide/User/Info/ptycho/neurips_demo_gpt4o_mini_l1_ptycho_script_1.m')" 
20240913-00:12:23 - [init] : Preparing paths.
[Warning: Applying custom scan flip: 1 1 1 ] 
[> In scans.positions.matlab_pos (line 14)
In scans.read_positions (line 20)
In core.initialize_ptycho (line 107)
In core.ptycho_recons/ptycho_call (line 172)
In core.ptycho_recons (line 289)
In ptycho_recon (line 576)
In ptycho_recon_pear (line 114)
In neurips_demo_gpt4o_mini_l1_ptycho_script_1 (line 44)
In run (line 99)
In driver (line 11)] 
[init] : Preparing initial guess.
[init] : Preparing data using matlab APS data preparation.
[init] : Finished data preparation and initialization.
[ptycho] : Reconstructing S00031 
[ptycho] : Calling engine GPU_MS
[GPU/CPU_MS-engine] : Intermediate results will be saved in the directory below every 50 iterations.
[GPU/CPU_MS-engine] : /Modified/to/Hide/User/Info/
[Warning: Applying custom data flip: 0 1 0 ] 
[ptycho] : Elapsed time for engine GPU_MS: 425.7 s
Elapsed time is 438.322446 seconds.

20240913-00:12:23 - PEAR: Your command execution is successful :) Please take a look and give me some feedback.
20240913-00:12:23 - Agent: QualityCollector
20240913-00:12:23 - PEAR: Please check your reconstructions and answer some questions:
20240913-00:12:24 - PEAR: Did the reconstruction converge?
20240913-00:12:44 - User: yes
20240913-00:12:44 - PEAR: Do you see any grid artifacts in the object?
20240913-00:12:47 - User: no
20240913-00:12:48 - PEAR: Is the initial probe accurate enough?
20240913-00:12:50 - User: yes
20240913-00:12:50 - PEAR: Do you see any structures in the last probe mode?
20240913-00:12:53 - User: yes
20240913-00:12:53 - PEAR: Is there anything else you want to tell me about the results?
20240913-00:13:01 - User: nothing at this moment
20240913-00:13:01 - PEAR: Thank you for your feedback!
20240913-00:13:01 - Agent: UpdatesRecommender
20240913-00:13:01 - PEAR: Next I will suggest some changes to the parameters:
20240913-00:13:01 - Agent: _summarize_conversation
20240913-00:13:01 - PEAR: Summarizing previous conversation...
20240913-00:13:02 - - The reconstruction converged successfully.  
- There are no grid artifacts present in the object.  
- The initial probe is accurate enough.  
- Structures are visible in the last probe mode.  
- There is nothing else to report about the results at this moment.  
20240913-00:13:03 - Based on the CONVERSATION SUMMARY, the situation is as follows:

- The initial probe is accurate enough.
- The reconstruction has converged.
- The last probe mode has structures.

### Recommendation:
* Increase the number of probe modes by 3.
20240913-00:13:03 - Agent: ParamsUpdater
20240913-00:13:07 - PEAR: Here are the updated parameters that will be used to create the next reconstruction script
20240913-00:13:07 - {
    "data_directory": "/Modified/to/Hide/User/Info/",
    "scan_number": 31,
    "beam_energy": 300,
    "radius_bright_field": 23.25,
    "convergence_angle": 25,
    "size_of_diffraction_patterns": 128,
    "use_external_object": false,
    "initial_object_path": "",
    "use_external_probe": true,
    "initial_probe_file": "/Modified/to/Hide/User/Info/Niter1000.mat",
    "defocus": 0,
    "use_external_positions": false,
    "initial_position_file": "",
    "grid_scan_positions": true,
    "scan_step_size_x": 0.2546,
    "scan_step_size_y": 0.2546,
    "number_scan_points_x": 128,
    "number_scan_points_y": 128,
    "number_of_iterations": 50,
    "update_batch_size": 128,
    "number_of_probe_modes": 6,
    "position_correction": false,
    "multislice_ptycho": true,
    "object_thickness": 100,
    "number_of_layers": 10,
    "layer_regularization_coefficient": 0,
    "diff_pattern_blur": 1,
    "gpu_id": 1
}
20240913-00:13:07 - PEAR: Do you want to run another reconstruction? If so, please say "yes".
20240913-00:13:18 - User: yes
20240913-00:13:18 - ########## Reconstruction No.2: ##########
20240913-00:13:18 - Agent: ParamsConfirmer
20240913-00:13:18 - PEAR: Let me know any changes you would like to make. If it looks good to you, please say "LGTM".
20240913-00:13:21 - User: lgtm
20240913-00:13:21 - PEAR: Thank you for the feedback. I will create a reconstruction script using these parameters.
20240913-00:13:21 - Agent: ScriptGenerator
20240913-00:13:21 - PEAR: Generating a reconstruction script...
20240913-00:13:26 - PEAR: The reconstruction script has been generated at /Modified/to/Hide/User/Info/ptycho/neurips_demo_gpt4o_mini_l1_ptycho_script_2.m
20240913-00:13:26 - Agent: ScriptRunner
20240913-00:13:26 - PEAR: Running ptychographic reconstruction...
20240913-00:13:26 - PEAR: Executing script using /usr/local/bin/matlab on Modified-to-Hide-User-Info...
20240913-00:13:26 - PEAR: The MATLAB command used for reconstruction: /usr/local/bin/matlab -batch "driver('/Modified/to/Hide/User/Info/ptycho/neurips_demo_gpt4o_mini_l1_ptycho_script_2.m')" 
20240913-00:13:26 - PEAR: Executing command /usr/local/bin/matlab -batch "driver('/Modified/to/Hide/User/Info/ptycho/neurips_demo_gpt4o_mini_l1_ptycho_script_2.m')" 
20240913-00:25:12 - [init] : Preparing paths.
[Warning: Applying custom scan flip: 1 1 1 ] 
[> In scans.positions.matlab_pos (line 14)
In scans.read_positions (line 20)
In core.initialize_ptycho (line 107)
In core.ptycho_recons/ptycho_call (line 172)
In core.ptycho_recons (line 289)
In ptycho_recon (line 576)
In ptycho_recon_pear (line 114)
In neurips_demo_gpt4o_mini_l1_ptycho_script_2 (line 44)
In run (line 99)
In driver (line 11)] 
[init] : Preparing initial guess.
[init] : Preparing data using matlab APS data preparation.
[init] : Finished data preparation and initialization.
[ptycho] : Reconstructing S00031 
[ptycho] : Calling engine GPU_MS
[GPU/CPU_MS-engine] : Intermediate results will be saved in the directory below every 50 iterations.
[GPU/CPU_MS-engine] : /Modified/to/Hide/User/Info/
[Warning: Applying custom data flip: 0 1 0 ] 
[ptycho] : Elapsed time for engine GPU_MS: 684.4 s
Elapsed time is 697.096650 seconds.

20240913-00:25:12 - PEAR: Your command execution is successful :) Please take a look and give me some feedback.
20240913-00:25:12 - Agent: QualityCollector
20240913-00:25:12 - PEAR: Please check your reconstructions and answer some questions:
20240913-00:25:13 - PEAR: Did the reconstruction converge?
20240913-00:25:27 - User: yes
20240913-00:25:27 - PEAR: Do you see any grid artifacts in the object?
20240913-00:25:30 - User: no
20240913-00:25:30 - PEAR: Is the initial probe accurate enough?
20240913-00:25:34 - User: yes
20240913-00:25:34 - PEAR: Do you see any structures in the last probe mode?
20240913-00:25:42 - User: not anymore
20240913-00:25:43 - PEAR: Is there anything else you want to tell me about the results?
20240913-00:26:19 - User: each layer in the reconstructed object seems to have random features that are not real
20240913-00:26:20 - PEAR: Thank you for your feedback!
20240913-00:26:20 - Agent: UpdatesRecommender
20240913-00:26:20 - PEAR: Next I will suggest some changes to the parameters:
20240913-00:26:20 - Agent: _summarize_conversation
20240913-00:26:20 - PEAR: Summarizing previous conversation...
20240913-00:26:21 - - The reconstruction converged successfully.  
- There are no grid artifacts present in the object.  
- The initial probe is accurate enough.  
- There are no visible structures in the last probe mode.  
- Each layer in the reconstructed object seems to have random features that are not real.  
20240913-00:26:21 - * Increase the layer regularization by 0.3.
20240913-00:26:21 - Agent: ParamsUpdater
20240913-00:26:25 - PEAR: Here are the updated parameters that will be used to create the next reconstruction script
20240913-00:26:25 - {
    "data_directory": "/Modified/to/Hide/User/Info/",
    "scan_number": 31,
    "beam_energy": 300,
    "radius_bright_field": 23.25,
    "convergence_angle": 25,
    "size_of_diffraction_patterns": 128,
    "use_external_object": false,
    "initial_object_path": "",
    "use_external_probe": true,
    "initial_probe_file": "/Modified/to/Hide/User/Info/Niter1000.mat",
    "defocus": 0,
    "use_external_positions": false,
    "initial_position_file": "",
    "grid_scan_positions": true,
    "scan_step_size_x": 0.2546,
    "scan_step_size_y": 0.2546,
    "number_scan_points_x": 128,
    "number_scan_points_y": 128,
    "number_of_iterations": 100,
    "update_batch_size": 100,
    "number_of_probe_modes": 5,
    "position_correction": false,
    "multislice_ptycho": true,
    "object_thickness": 100,
    "number_of_layers": 10,
    "layer_regularization_coefficient": 0.3,
    "diff_pattern_blur": 1,
    "gpu_id": 1
}
20240913-00:26:25 - PEAR: Do you want to run another reconstruction? If so, please say "yes".
20240913-00:27:10 - User: yes
20240913-00:27:10 - ########## Reconstruction No.3: ##########
20240913-00:27:10 - Agent: ParamsConfirmer
20240913-00:27:10 - PEAR: Let me know any changes you would like to make. If it looks good to you, please say "LGTM".
20240913-00:27:23 - User: lgtm
20240913-00:27:23 - PEAR: Thank you for the feedback. I will create a reconstruction script using these parameters.
20240913-00:27:23 - Agent: ScriptGenerator
20240913-00:27:23 - PEAR: Generating a reconstruction script...
20240913-00:27:29 - PEAR: The reconstruction script has been generated at /Modified/to/Hide/User/Info/ptycho/neurips_demo_gpt4o_mini_l1_ptycho_script_3.m
20240913-00:27:29 - Agent: ScriptRunner
20240913-00:27:29 - PEAR: Running ptychographic reconstruction...
20240913-00:27:29 - PEAR: Executing script using /usr/local/bin/matlab on Modified-to-Hide-User-Info...
20240913-00:27:29 - PEAR: The MATLAB command used for reconstruction: /usr/local/bin/matlab -batch "driver('/Modified/to/Hide/User/Info/ptycho/neurips_demo_gpt4o_mini_l1_ptycho_script_3.m')" 
20240913-00:27:29 - PEAR: Executing command /usr/local/bin/matlab -batch "driver('/Modified/to/Hide/User/Info/ptycho/neurips_demo_gpt4o_mini_l1_ptycho_script_3.m')" 
20240913-00:49:23 - [init] : Preparing paths.
[Warning: Applying custom scan flip: 1 1 1 ] 
[> In scans.positions.matlab_pos (line 14)
In scans.read_positions (line 20)
In core.initialize_ptycho (line 107)
In core.ptycho_recons/ptycho_call (line 172)
In core.ptycho_recons (line 289)
In ptycho_recon (line 576)
In ptycho_recon_pear (line 114)
In neurips_demo_gpt4o_mini_l1_ptycho_script_3 (line 44)
In run (line 99)
In driver (line 11)] 
[init] : Preparing initial guess.
[init] : Preparing data using matlab APS data preparation.
[init] : Finished data preparation and initialization.
[ptycho] : Reconstructing S00031 
[ptycho] : Calling engine GPU_MS
[GPU/CPU_MS-engine] : Intermediate results will be saved in the directory below every 50 iterations.
[GPU/CPU_MS-engine] : /Modified/to/Hide/User/Info/
[Warning: Applying custom data flip: 0 1 0 ] 
[ptycho] : Elapsed time for engine GPU_MS: 1293.5 s
Elapsed time is 1306.163743 seconds.

20240913-00:49:23 - PEAR: Your command execution is successful :) Please take a look and give me some feedback.
20240913-00:49:23 - Agent: QualityCollector
20240913-00:49:23 - PEAR: Please check your reconstructions and answer some questions:
20240913-00:49:23 - PEAR: Did the reconstruction converge?
20240913-00:56:04 - User: yes
20240913-00:56:05 - PEAR: Do you see any grid artifacts in the object?
20240913-00:56:15 - User: no
20240913-00:56:15 - PEAR: Is the initial probe accurate enough?
20240913-00:56:18 - User: yes
20240913-00:56:18 - PEAR: Do you see any structures in the last probe mode?
20240913-00:56:28 - User: no
20240913-00:56:28 - PEAR: Is there anything else you want to tell me about the results?
20240913-00:56:42 - User: the atoms still look a bit blurred
20240913-00:56:43 - PEAR: Thank you for your feedback!
20240913-00:56:43 - Agent: UpdatesRecommender
20240913-00:56:43 - PEAR: Next I will suggest some changes to the parameters:
20240913-00:56:43 - Agent: _summarize_conversation
20240913-00:56:43 - PEAR: Summarizing previous conversation...
20240913-00:56:44 - - The reconstruction converged successfully.  
- There are no grid artifacts present in the object.  
- The initial probe is accurate enough.  
- There are no visible structures in the last probe mode.  
- The atoms still appear a bit blurred.  
20240913-00:56:44 - * Increase the standard deviation of the gaussian kernel for detector blurring by 0.5.
20240913-00:56:44 - Agent: ParamsUpdater
20240913-00:56:47 - PEAR: Here are the updated parameters that will be used to create the next reconstruction script
20240913-00:56:47 - {
    "data_directory": "/Modified/to/Hide/User/Info/",
    "scan_number": 31,
    "beam_energy": 300,
    "radius_bright_field": 23.25,
    "convergence_angle": 25,
    "size_of_diffraction_patterns": 128,
    "use_external_object": false,
    "initial_object_path": "",
    "use_external_probe": true,
    "initial_probe_file": "/Modified/to/Hide/User/Info/Niter1000.mat",
    "defocus": 0,
    "use_external_positions": false,
    "initial_position_file": "",
    "grid_scan_positions": true,
    "scan_step_size_x": 0.2546,
    "scan_step_size_y": 0.2546,
    "number_scan_points_x": 128,
    "number_scan_points_y": 128,
    "number_of_iterations": 100,
    "update_batch_size": 100,
    "number_of_probe_modes": 5,
    "position_correction": false,
    "multislice_ptycho": true,
    "object_thickness": 100,
    "number_of_layers": 10,
    "layer_regularization_coefficient": 0.3,
    "diff_pattern_blur": 1.5,
    "gpu_id": 1
}
20240913-00:56:47 - PEAR: Do you want to run another reconstruction? If so, please say "yes".
20240913-00:57:20 - User: yes
20240913-00:57:20 - ########## Reconstruction No.4: ##########
20240913-00:57:20 - Agent: ParamsConfirmer
20240913-00:57:20 - PEAR: Let me know any changes you would like to make. If it looks good to you, please say "LGTM".
20240913-00:57:34 - User: set the update batch size to 256
20240913-00:57:38 - PEAR: Here are the updated parameters:
20240913-00:57:38 - {
    "data_directory": "/Modified/to/Hide/User/Info/",
    "scan_number": 31,
    "beam_energy": 300,
    "radius_bright_field": 23.25,
    "convergence_angle": 25,
    "size_of_diffraction_patterns": 128,
    "use_external_object": false,
    "initial_object_path": "",
    "use_external_probe": true,
    "initial_probe_file": "/Modified/to/Hide/User/Info/Niter1000.mat",
    "defocus": 0,
    "use_external_positions": false,
    "initial_position_file": "",
    "grid_scan_positions": true,
    "scan_step_size_x": 0.2546,
    "scan_step_size_y": 0.2546,
    "number_scan_points_x": 128,
    "number_scan_points_y": 128,
    "number_of_iterations": 100,
    "update_batch_size": 256,
    "number_of_probe_modes": 5,
    "position_correction": false,
    "multislice_ptycho": true,
    "object_thickness": 100,
    "number_of_layers": 10,
    "layer_regularization_coefficient": 0.3,
    "diff_pattern_blur": 1.5,
    "gpu_id": 1
}
20240913-00:57:38 - PEAR: Let me know any changes you would like to make. If it looks good to you, please say "LGTM".
20240913-00:57:50 - User: lgtm
20240913-00:57:50 - PEAR: Thank you for the feedback. I will create a reconstruction script using these parameters.
20240913-00:57:50 - Agent: ScriptGenerator
20240913-00:57:50 - PEAR: Generating a reconstruction script...
20240913-00:57:56 - PEAR: The reconstruction script has been generated at /Modified/to/Hide/User/Info/ptycho/neurips_demo_gpt4o_mini_l1_ptycho_script_4.m
20240913-00:57:56 - Agent: ScriptRunner
20240913-00:57:56 - PEAR: Running ptychographic reconstruction...
20240913-00:57:56 - PEAR: Executing script using /usr/local/bin/matlab on Modified-to-Hide-User-Info...
20240913-00:57:56 - PEAR: The MATLAB command used for reconstruction: /usr/local/bin/matlab -batch "driver('/Modified/to/Hide/User/Info/ptycho/neurips_demo_gpt4o_mini_l1_ptycho_script_4.m')" 
20240913-00:57:56 - PEAR: Executing command /usr/local/bin/matlab -batch "driver('/Modified/to/Hide/User/Info/ptycho/neurips_demo_gpt4o_mini_l1_ptycho_script_4.m')" 
20240913-01:17:09 - [init] : Preparing paths.
[Warning: Applying custom scan flip: 1 1 1 ] 
[> In scans.positions.matlab_pos (line 14)
In scans.read_positions (line 20)
In core.initialize_ptycho (line 107)
In core.ptycho_recons/ptycho_call (line 172)
In core.ptycho_recons (line 289)
In ptycho_recon (line 576)
In ptycho_recon_pear (line 114)
In neurips_demo_gpt4o_mini_l1_ptycho_script_4 (line 44)
In run (line 99)
In driver (line 11)] 
[init] : Preparing initial guess.
[init] : Preparing data using matlab APS data preparation.
[init] : Finished data preparation and initialization.
[ptycho] : Reconstructing S00031 
[ptycho] : Calling engine GPU_MS
[GPU/CPU_MS-engine] : Intermediate results will be saved in the directory below every 50 iterations.
[GPU/CPU_MS-engine] : /Modified/to/Hide/User/Info/
[Warning: Applying custom data flip: 0 1 0 ] 
[ptycho] : Elapsed time for engine GPU_MS: 1132.4 s
Elapsed time is 1145.022578 seconds.

20240913-01:17:09 - PEAR: Your command execution is successful :) Please take a look and give me some feedback.
20240913-01:17:09 - Agent: QualityCollector
20240913-01:17:09 - PEAR: Please check your reconstructions and answer some questions:
20240913-01:17:09 - PEAR: Did the reconstruction converge?
20240913-01:19:28 - User: yes
20240913-01:19:28 - PEAR: Do you see any grid artifacts in the object?
20240913-01:19:30 - User: no
20240913-01:19:31 - PEAR: Is the initial probe accurate enough?
20240913-01:19:36 - User: yes
20240913-01:19:36 - PEAR: Do you see any structures in the last probe mode?
20240913-01:19:39 - User: no
20240913-01:19:40 - PEAR: Is there anything else you want to tell me about the results?
20240913-01:19:42 - User: no
20240913-01:19:43 - PEAR: Thank you for your feedback!
20240913-01:19:43 - Agent: UpdatesRecommender
20240913-01:19:43 - PEAR: Next I will suggest some changes to the parameters:
20240913-01:19:43 - Agent: _summarize_conversation
20240913-01:19:43 - PEAR: Summarizing previous conversation...
20240913-01:19:44 - - The reconstruction converged successfully.  
- There are no grid artifacts present in the object.  
- The initial probe is accurate enough.  
- There are no structures visible in the last probe mode.  
- The user has no additional information to provide about the results.  
- All diagnosis questions have been asked.  
20240913-01:19:45 - * Increase the standard deviation of the gaussian kernel for detector blurring by 0.5.
20240913-01:19:45 - Agent: ParamsUpdater
20240913-01:19:48 - PEAR: Here are the updated parameters that will be used to create the next reconstruction script
20240913-01:19:48 - {
    "data_directory": "/Modified/to/Hide/User/Info/",
    "scan_number": 31,
    "beam_energy": 300,
    "radius_bright_field": 23.25,
    "convergence_angle": 25,
    "size_of_diffraction_patterns": 128,
    "use_external_object": false,
    "initial_object_path": "",
    "use_external_probe": true,
    "initial_probe_file": "/Modified/to/Hide/User/Info//Niter1000.mat",
    "defocus": 0,
    "use_external_positions": false,
    "initial_position_file": "",
    "grid_scan_positions": true,
    "scan_step_size_x": 0.2546,
    "scan_step_size_y": 0.2546,
    "number_scan_points_x": 128,
    "number_scan_points_y": 128,
    "number_of_iterations": 100,
    "update_batch_size": 256,
    "number_of_probe_modes": 5,
    "position_correction": false,
    "multislice_ptycho": true,
    "object_thickness": 100,
    "number_of_layers": 10,
    "layer_regularization_coefficient": 0.3,
    "diff_pattern_blur": 2.0,
    "gpu_id": 1
}
20240913-01:19:48 - PEAR: Do you want to run another reconstruction? If so, please say "yes".
20240913-01:20:20 - User: no
20240913-01:20:20 - PEAR: Reconstructions completed. Have a nice day!
\end{verbatim}

\end{document}